\documentclass[aps, prb, twocolumn, showpacs]{revtex4}

\usepackage{amsmath}
\usepackage{graphicx}

\newcommand{\espacio}{\qquad\quad}
\newcommand{\sign}{\operatorname{sign}}
\newcommand{\bra}{\left\langle}
\newcommand{\ket}{\right\rangle}

\newcommand{\zb}{\bar{z}}

\newcommand{\cd}{c^{\dagger}}
\renewcommand{\O}{\mathcal{O}}
\newcommand{\T}{\mathcal{T}}

\begin{document}

\title{Correlation functions for 1d interacting fermions with spin-orbit coupling}
\author{An\'{\i}bal Iucci}
\email{iucci@fisica.unlp.edu.ar}

\affiliation{Instituto de F\'{\i}sica La Plata. Departamento de
F\'{\i}sica, Facultad de Ciencias Exactas, Universidad Nacional de
La Plata. CC 67, 1900 La Plata, Argentina.\\ Consejo Nacional de
Investigaciones Cient\'{\i}ficas y T\'ecnicas, Argentina.}

\begin{abstract}
We compute correlation functions for one-dimensional electron
systems which spin and charge degrees of freedom are coupled
through spin-orbit coupling. Charge density waves, spin density
waves, singlet- triplet- superconducting fluctuations are studied.
We show that the spin-orbit interaction modify the exponents and
the phase diagram of the system, changing the dominant
fluctuations and making new susceptibilities diverge for low
temperature.
\end{abstract}



\maketitle

In past two decades there have been an intense effort in studying
quasi-one-dimensional electron systems (Q1DES). This interest has
its origin in the simplicity of the models which describe them,
and at the same time, in the possibility of making contact with
experiments. Examples of these Q1DES of recent construction are
carbon nanotubes\cite{bockrath99}, conducting
polymers\cite{heeger88} and semiconductor
heterostructures\cite{tarucha95}. From the theoretical point of
view the simplest formulation of a Q1DES is given by the
Tomonaga-Luttinger \cite{tomonaga50, reviews} model which
describes the major qualitative features of the interacting Q1DES
such as the spin-charge separation and the non universal exponents
in the decay law of correlation functions.

In realistic situations the electrons are moving in electric
fields inside the materials: microscopic and macroscopic ones, the
latter responsible of confining the electrons to a reduced region
of space. As a consequence it appears a magnetic field in the rest
frame of the electron which couples with its intrinsic magnetic
moment and breaks the spin-rotation SU(2) symmetry. This is known
as spin-orbit (SO) interaction (or spin-orbit coupling). Despite
its relativistic origin, this interaction has an important effect
on existing two-dimensional electron gases (2DEG) such as
GaAs/AlGaAs\cite{Hassenkam97, Miller03} and
InGaAs/InAlAs\cite{Luo90, Nitta97} heterostructures. It is
responsible for the modification of their band structure by
lifting the spin degeneracy and for positive magnetoresistance
effects, known as weak antilocalization\cite{Bergmann84}

In Q1DES there exists an additional potential, responsible of
confining the electrons in a narrow channel patterned in 2DEG
heterostructures\cite{Meirav89}. Although as far as we know there
is no experimental evidence or measures of the strength of the SO
coupling resulting from such a potential, theoretical work
indicates that it affects quantitatively the splitting energy
behavior as a function of the wave vector $k$, and produces an
asymmetric deformation of each spin branch, i.e. the Fermi
velocities take different values for different directions of
motion\cite{moroz99}.

In Ref. \onlinecite{moroz00} the following Hamiltonian was
proposed as a model for a Q1DES with SO coupling

\begin{equation}\label{Hferm}
H=H_0+H_\text{int}
\end{equation}
where the noninteracting Hamiltonian is

\begin{multline}\label{HfermFree}
H_0=v_1\sum_k \left[(k+k_1)\cd_{kR\uparrow}c_{kR\uparrow}
-(k-k_1)\cd_{kL\downarrow}c_{kL\downarrow}\right]\\
+v_2\sum_k \left[(k+k_2)\cd_{kR\downarrow}c_{kR\downarrow}
-(k-k_2)\cd_{kL\uparrow}c_{kL\uparrow}\right].
\end{multline}
It consists of a modified Tomonaga-Luttinger model that takes into
account the asymmetry in the spectrum for each spin branch making
$v_1\neq v_2$ (and $k_1\neq k_2)$. $\cd_{krs}$ creates a
right-going (r=+1) or left-going (r=-1) electron. The interacting
Hamiltonian describes forward scattering electron-electron
interactions and has a standard form\cite{reviews}. Umklapp and
backscattering terms are irrelevant if we are far from half
0filling in the former case, and we restrict to repulsive
interactions in the latter.

In this article we compute correlation functions for charge
density wave (CDW), spin density wave (SDW) $4k_F$ charge density
wave ($4k_F$) and singlet- and triplet- superconductivity (SS and
TS) operators for the model presented above. The correlation
functions for these operators are well-known in the case of zero
spin-orbit coupling\cite{solyom79, emery79, reviews}, including
logarithmic correction factors\cite{giamarchi89, voit92} and time
and temperature dependence\cite{emery79}. We extend these
calculations to the case in which SO interaction are present, and
study how the exponents of their algebraic decay are modified. We
find interesting modifications of the phase diagram of the system
when SO interactions are present. For certain regions of the
parameter space, SO coupling changes the dominant fluctuations,
and makes new susceptibilities diverge for low temperature.

The Hamiltonian (\ref{Hferm}) can be studied by the use of
bosonization technique \cite{reviews, haldane81} as in Ref.
\onlinecite{moroz00}. For convenience we shall define an average
velocity $v_0=(v_1+v_2)/2$ and the difference $\delta v=v_2-v_1$,
and the same for the Fermi momentum $k_0=(k_1+k_2)/2$ and $\delta
k=k_2-k_1$. If we introduce the usual phase fields
$\phi_\rho$($\phi_\sigma$) for charge (spin) degrees of freedom
and the dual field $\Pi_\rho$($\Pi_\sigma$) the Hamiltonian can be
represented as

\begin{equation}
\begin{split}\label{Hbos}
H=&\frac{v_\rho}{2}\int dx
\left[\frac{1}{K_\rho}(\partial_x\phi_\rho)^2+K_\rho\Pi_\rho^2\right]\\
+&\frac{v_\sigma}{2}\int dx
\left[\frac{1}{K_\sigma}(\partial_x\phi_\sigma)^2+K_\sigma\Pi_\sigma^2\right]\\
+& \delta v\int dx
\left[(\partial_x\phi_\rho)\Pi_\sigma+(\partial_x\phi_\sigma)\Pi_\rho\right].
\end{split}
\end{equation}
$v_{\rho,\sigma}$ are the propagation velocities of the spin and
charge collective modes of the decoupled model ($\delta v=0$), and
$K_{\rho,\sigma}$ are the stiffness constants. The spin-orbit
interaction appears as an effect that breaks the spin charge
separation as reveals the presence of the third term in the last
equation. Nevertheless the Hamiltonian (\ref{Hbos}) can still be
diagonalized in terms of two new phase fields which contains a
mixture of spin and charge degrees of freedom. The propagation
velocities of these collective modes are

\begin{multline}
v_\pm^2=\frac{v_\sigma^2+v_\rho^2}{2}+\delta v^2\\
\pm\sqrt{\left(\frac{v_\rho^2-v_\sigma^2}{2}\right)^2+\delta
v^2\left[v_\sigma^2+v_\rho^2+v_\rho
v_\sigma\left(\frac{K_\rho}{K_\sigma}+\frac{K_\sigma}{K_\rho}\right)\right]}.
\end{multline}
As $\delta v\rightarrow 0$, $v_+\rightarrow\max(v_\rho,v_\sigma)$
and $v_-\rightarrow\min(v_\rho,v_\sigma)$. As $\delta v$
increases, $v_-$ decreases until it vanishes at the points

\begin{align}
\delta v_\rho^2=&v_\rho v_\sigma\frac{K_\sigma}{K_\rho}\\
\delta v_\sigma^2=&v_\rho v_\sigma\frac{K_\rho}{K_\sigma}
\end{align}

At these points, the freezing of the lower bosonic mode is
accompanied by a divergence in the charge and spin response
functions. The static charge compressibility $\kappa$ diverges at
$\delta v=\delta v_\rho$, and at $\delta v=\delta v_\sigma$ occurs
a divergence of the static spin susceptibility $\chi$. They behave
as

\begin{align}\label{instabilities1}
\kappa=&\kappa_0\left[1-\frac{\delta v}{\delta
v_\rho}\right]^{-1},\espacio\kappa_0=\frac{2K_\rho}{\pi v_\rho}\\
\chi=&\chi_0\left[1-\frac{\delta v}{\delta
v_\sigma}\right]^{-1},\espacio\chi_0=\frac{2K_\sigma}{\pi
v_\sigma}\label{instabilities2}
\end{align}
where $\kappa_0$ and $\chi_0$ are the values of $\kappa$ and
$\chi$ in absence of SO coupling. Beyond these points the
susceptibilities becomes negatives. This behavior of the static
response functions together with the vanishing of the collective
modes velocity indicates that the system becomes
unstable\cite{reviews, drut02} and undergoes a first order phase
transition\cite{voit92}. For $K_\rho>K_\sigma$, $\delta v_\rho$
turns out to be lower than $\delta v_\sigma$ and as $\delta v$
grows from the zero value the physical divergence takes place in
the charge compressibility. This instability is known as phase
separation and has been shown to occur in the extended Hubbard
model\cite{penc} and in the t-J model\cite{t-J}. In the case that
$K_\rho<K_\sigma$, the instability takes place in the spin
subsystem and is related to the so called metamagnetic transition,
observed for instance in the quasi-one-dimensional compound
$\text{Ba}_3\text{Cu}_2\text{O}_4\text{Cl}2$\cite{metaExp}. It
also arises in the phase diagram of the XXZ model with
next-to-nearest neighbors\cite{XXZ}. In presence of a chemical
potential (magnetic field), the region where $\kappa$ ($\chi$) is
negative is associated with the coexistence of two phases with
different hole concentration (magnetization). The divergence of
$\kappa$ was found in other models with asymmetric
dispersion\cite{fernandez02}.

Let us now focus our attention on the correlation functions. Our
interest in this work is to obtain their space-time and
temperature $T=1/\beta$ behavior. The operators for CDW, SDW,
4$k_F$, SS and TS fluctuations in their bosonized form are

\begin{align}
\O_{\text{CDW}}&=\frac{2}{\pi
a}\cos(2k_0x+\sqrt{2\pi}\phi_\rho)\cos\sqrt{2\pi}\phi_\sigma\\
\O_{4k_F}&=\frac{1}{(\pi a)^2}\cos(4k_0x+\sqrt{8\pi}\phi_\rho)\\
\O_{\text{SDW},x}&=\frac{2}{\pi a}
\cos(2k_0x+\sqrt{2\pi}\phi_\rho)\cos(\delta k
x+\sqrt{2\pi}\theta_\sigma)\\
\O_{\text{SDW},y}&=\frac{2}{\pi a}
\cos(2k_0x+\sqrt{2\pi}\phi_\rho)\sin(\delta k
x+\sqrt{2\pi}\theta_\sigma)\\
\O_{\text{SDW},z}&=\frac{2}{\pi a}
\sin(2k_0x+\sqrt{2\pi}\phi_\rho)\sin\sqrt{2\pi}\phi_\sigma\\
\O_{\text{SS}}&=\frac{-i}{\sqrt{2}\pi a}
e^{-i\sqrt{2\pi}\theta_\rho}\sin\sqrt{2\pi}\phi_\sigma\\
\O_{\text{TS},0}&=\frac{1}{\sqrt{2}\pi a}
e^{-i\sqrt{2\pi}\theta_\rho}\cos\sqrt{2\pi}\phi_\sigma\\
\O_{\text{TS},\pm1}&=\frac{1}{2\pi a}e^{\pm i\delta k
x}e^{-i\sqrt{2\pi}(\theta_\rho\pm\theta_\sigma)}
\end{align}
where $a$ is a short distance cutoff and $\theta_\nu$ is related
to the conjugated field $\Pi_\nu$ by the relation
$\Pi_\lambda=\partial_x\theta_\lambda$.

The correlation functions are defined as

\begin{equation}
R_i(x,\tau;\beta)=\bra\T_\tau\O_i(x,\tau)\O^\dagger_i(0,0)\ket.
\end{equation}
where $\T_\tau$ is the (imaginary) time-ordering operator. These
objects were calculated in the path integral framework within the
Matsubara imaginary time formalism and the results are:

\begin{widetext} 

\begin{multline}
R_{\text{CDW}}(x,\tau;\beta)=R_{\text{SDW},z}(x,\tau;\beta)=\\\frac{\cos{2k_0x}}{2(\pi{a})^2}
\left(z_+ \zb_+\right)^
{-(K_\rho\nu_+^\rho+K_\sigma\nu_+^\sigma)/2}
\left(z_-\zb_-\right)^
{-(K_\rho\nu_-^\rho+K_\sigma\nu_-^\sigma)/2}
\left[\left(\frac{\zb_+z_-}{z_+\zb_-}\right)^{H\sign(x\tau)}+h.c.\right]
\end{multline}

\begin{multline}\label{SDWxy}
R_{\text{SDW},xy}(x,\tau;\beta)=\frac{\cos{2k_1x}}{2(\pi{a})^2}
\left(z_+\zb_+\right)^{-(K_\rho\nu_+^\rho+\mu_+^\sigma/K_\sigma)/2-\theta_+^\sigma}
\left(z_-\zb_-\right)^{-(K_\rho\nu_-^\rho+\mu_-^\sigma/K_\sigma)/2-\theta_-^\sigma}\\+
\frac{\cos{2k_2x}}{2(\pi{a})^2}
\left(z_+\zb_+\right)^{-(K_\rho\nu_+^\rho+\mu_+^\sigma/K_\sigma)/2+\theta_+^\sigma}
\left(z_-\zb_-\right)^{-(K_\rho\nu_-^\rho+\mu_-^\sigma/K_\sigma)/2+\theta_-^\sigma}
\end{multline}

\begin{equation}
R_{4k_F}(x,\tau;\beta)=\frac{\cos{4k_0x}}{2(\pi{a})^4}
\left(z_+\zb_+\right)^{-2K_\rho\nu_+^\rho}
\left(z_-\zb_-\right)^{-2K_\rho\nu_-^\rho}
\end{equation}

\begin{equation}
R_{\text{SS}}(x,\tau;\beta)=R_{\text{TS},0}(x,\tau;\beta)=\frac{1}{2(2\pi{a})^2}
\left(z_+\zb_+\right)^{-(\mu_+^\rho/K_\rho+K_\sigma\nu_+^\sigma)/2+\theta_+^\rho}
\left(z_-\zb_-\right)^{-(\mu_-^\rho/K_\rho+K_\sigma\nu_-^\sigma)/2+\theta_-^\rho}+
(\theta_\pm^\rho\rightarrow -\theta_\pm^\rho)
\end{equation}

\begin{equation}
R_{\text{TS},\pm1}(x,\tau;\beta)=\frac{e^{\pm i\delta
kx}}{(2\pi{a})^2}
\left(z_+\zb_+\right)^{-(\mu_+^\rho/K_\rho+\mu_+^\sigma/K_\sigma)/2}
\left(z_-\zb_-\right)^{-(\mu_-^\rho/K_\rho+\mu_-^\sigma/K_\sigma)/2}
\left(\frac{\zb_+z_-}{z_+\zb_-}\right)^{\pm G\sign(x\tau)}
\end{equation}

\end{widetext}
where

\begin{align}
z_\pm=&\frac{\sin\frac{\pi}{v_\pm\beta}(v_\pm|\tau|+\epsilon+ix)}
{\sin{\frac{\pi\epsilon}{v_\pm\beta}}}\\
\zb_\pm=&\frac{\sin\frac{\pi}{v_\pm\beta}(v_\pm|\tau|+\epsilon-ix)}
{\sin{\frac{\pi\epsilon}{v_\pm\beta}}}
\end{align}
and the exponents depend on the the stiffness constants multiplied
by the factors that include mode velocities dependence. They are
given by

\begin{align}
\nu_\pm^\lambda=&\pm\frac{v_\lambda}{v_\pm}\frac{v^2_\pm-v^2_{-\lambda}\left(1-\delta
v^2/\delta v_{-\lambda}^2\right)}{v_+^2-v_-^2}\\
\mu_\pm^\lambda=&\pm\frac{v_\lambda}{v_\pm}\frac{v^2_\pm-v^2_{-\lambda}\left(1-\delta
v^2/\delta v_\lambda^2\right)}{v_+^2-v_-^2}\\
\theta_\pm^\lambda=&\pm\frac{\delta
v}{v_\pm}\frac{v^2_\pm-\left(\delta v_\lambda^2- \delta
v^2\right)}{v_+^2-v_-^2}\label{exptheta}
\end{align}
with $\lambda=\rho,\sigma$, and

\begin{align}
H=&\delta v\frac{K_\rho v_\rho+K_\sigma v_\sigma}{v_+^2-v_-^2}\\
G=&\delta v\frac{v_\rho/K_\rho + v_\sigma/K_\sigma
}{v_+^2-v_-^2}\label{expG}.
\end{align}
$\nu_\pm^\lambda$ and $\mu_\pm^\lambda$ are positive, and
$\theta_\pm^\lambda$, $G$ and $H$ has the same sign as $\delta v$.

In the model with zero SO coupling the SU(2) symmetry can be
restored by imposing the constraint $K_\sigma=1$, which emerges
naturally if the model under study is the continuum limit of a
lattice model with only charge density interactions. In this case
this is not possible; the SU(2) symmetry keeps broken even for
$K_\sigma=1$ as reveals the differences in the decays between SDW
operators correlation functions in the $z$ direction and in the
$x,y$ directions. As in the zero SO case, correlation functions
for SDW operators in the $z$ direction and CDW operators are
equal, and the same happen with TS,0 and SS operators. This
degeneracy is broken by logarithmic corrections that arises if
irrelevant backscattering or umklapp terms are
included\cite{giamarchi89}.

An interesting point to observe is the appearance of two terms in
the $\text{SDW},xy$ correlation functions (Eq. \ref{SDWxy}) where
the modulations have different frequencies and decay with
different exponents. As $\theta^\lambda_\pm$ has the same sign as
$\delta v$ (see Eq. \ref{exptheta} and the comment below Eq.
\ref{expG}) for $v_2>v_1$ ($v_2<v_1$) the dominant term is the one
with frequency $k_2$ ($k_1$). In other words the biggest frequency
dominates. Also $R_{\text{TS},\pm1}$ becomes oscillating.

Up to here we have obtained very general formulae for space-time
and temperature dependent correlation functions for the model
under analysis. We can gain physical insight by observing the
algebraic decay of the instantaneous correlation functions at zero
temperature, and studying how the exponents get modified from the
zero SO case. The functions behave as

\begin{equation}
R_i(x)\sim |x|^{-2+\alpha_i}.
\end{equation}
The exponents $\alpha_i'\text{s}$ determine the divergence of the
corresponding Fourier space susceptibility as $T\rightarrow 0$,
$\chi_i(T)\sim T^{-\alpha_i}$\cite{reviews}. This makes these
instabilities of a completely different nature than the ones
described in equations (\ref{instabilities1}) and
(\ref{instabilities2}). The expressions obtained for the
$\alpha_i$ are

\begin{align}
\alpha_{\text{CDW}}=\alpha_{\text{SDW},z}=&2-K_\rho\nu^\rho-K_\sigma\nu^\sigma\\
\alpha_{\text{SDW},x}=\alpha_{\text{SDW},y}=&2(1+|\theta^\sigma|)-K_\rho\nu^\rho-\mu^\sigma/K_\sigma\\
\alpha_{\text{SS}}=\alpha_{\text{TS},0}=&2(1+|\theta^\rho|)-\mu^\rho/K_\rho-K_\sigma\nu^\sigma\\
\alpha_{\text{TS},\pm 1}=&2-\mu^\rho/K_\rho-\mu^\sigma/K_\sigma.
\end{align}
These are the new exponents, which retain the same structure as in
the zero SO coupling, but modified by the factors

\begin{gather}
\mu^\lambda=\mu_+^\lambda+\mu_-^\lambda\\
\nu^\lambda=\nu_+^\lambda+\nu_-^\lambda\\
\theta^\lambda=\theta_+^\lambda+\theta_-^\lambda.
\end{gather}
When $\delta v\rightarrow 0$, $\theta^\lambda\rightarrow 0$ and
$\mu^\lambda,\nu^\lambda\rightarrow 1$, so we reproduce the right
results for the zero SO case.

\begin{figure}
\includegraphics{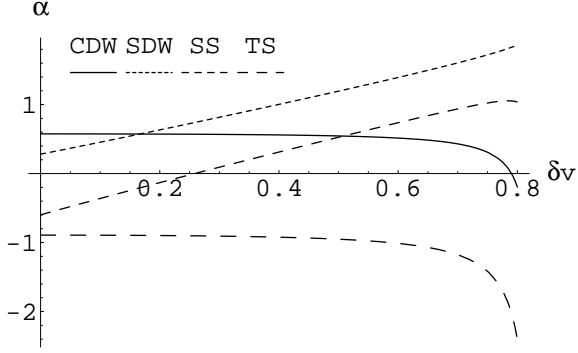}
\caption{\label{exponents} Behavior of the exponents
$\alpha_i'\text{s}$ as a function of $\delta v$ (in units of
$v_0$). For $v_\rho=1.2 v_0$, $v_\sigma=0.8 v_0$, $K_\rho=0.6$ and
$K_\sigma=0.85$. For $\delta v\gtrsim 0.16$ $\text{SDW},xy$
fluctuations get dominant, and for $\delta v\gtrsim 0.25$
$\alpha_{\text{SS}}$ gets positive, and $\chi_{\text{SS}}$
divergent for $T\rightarrow 0$.}
\end{figure}

For finite SO coupling, $\delta v$ appears as a parameter which
plays a role in determining the slowest decaying correlation
function, and which are the divergent susceptibilities. In Fig.
\ref{exponents} we observe as an example, the behavior of the
exponents as a function of $\delta v$ for $v_\rho=1.2 v_0$,
$v_\sigma=0.8 v_0$, $K_\rho=0.6$ and $K_\sigma=0.85$. For $\delta
v$ small, CDW fluctuations are dominant, but for $\delta v\gtrsim
0.16 v_0$ the $\text{SDW},xy$ correlations decay slower. For small
$\delta v$, CDW and SDW are the only diverging susceptibilities
for $T\rightarrow 0$, but for $\delta v\gtrsim 0.25 v_0$,
$\alpha_{\text{SS}}$ becomes positive, and $\chi_{\text{SS}}$
divergent for $T\rightarrow 0$. Calculations of the electron band
structure modified by SO coupling show that these values of
$\delta v$ should correspond to typical Q1DES\cite{moroz99}.

\begin{figure}
\includegraphics{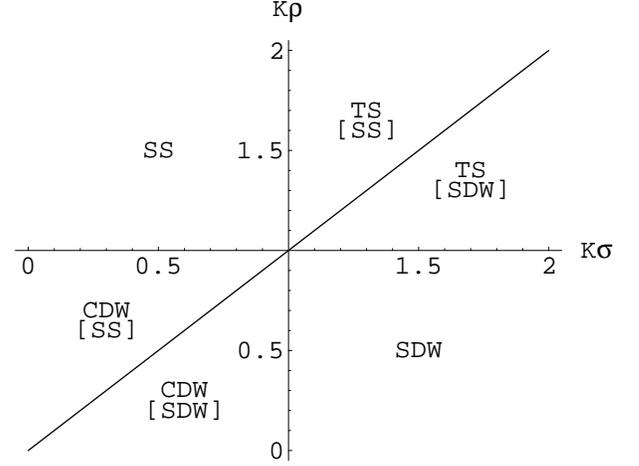}
\caption{\label{phaseDiag} Phase diagram in $K_\rho-K_\sigma$
space. The phase in brackets is the subdominant one, which becomes
dominant for strong enough SO coupling.}
\end{figure}

\begin{figure}
\includegraphics{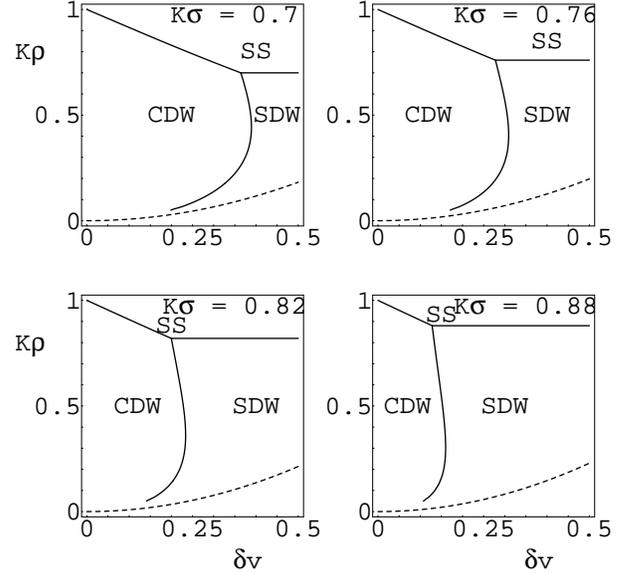}
\caption{\label{crossSections} Phase diagram in $K_\rho-\delta v$
space for $v_\rho=1.2 v_0$, $v_\sigma=0.8 v_0$ and different
values of $K_\sigma$. $\delta v>\delta v_\sigma$ below the doted
line and metamagnetism occurs. }
\end{figure}

A careful analysis of the exponents allows us to construct a phase
diagram in $K_\rho-K_\sigma$ space (Fig. \ref{phaseDiag}). In each
region we indicate the dominant fluctuation for small $\delta v$,
and in brackets, the dominant one for stronger $\delta v$. Other
subdominant fluctuations are not indicated. Cross sections of the
phase diagram are shown in Fig. \ref{crossSections}. In this plot
the $K_\rho-\delta v$ space can be observed for $K_\rho<1$ and
different values of $K_\sigma$. For small $\delta v$ CDW
fluctuations dominate and for stronger $\delta v$ the system can
be either in the SDW or in the SS phase depending on the values of
$K_\rho$ and $K_\sigma$. In the region below the doted line,
$\delta v>\delta v_\sigma$, the static spin susceptibility becomes
negative and metamagnetism takes place

In conclusion, we have computed correlation functions for a model
of one-dimensional correlated electrons with SO coupling. This
coupling destroys the spin charge separation as was shown in Ref.
\onlinecite{moroz00}, and modifies the exponents of correlations
decay. As a consequence the phase diagram gets modified. For
strong enough SO coupling, it changes the dominant fluctuation,
and makes new susceptibilities diverge for $T\rightarrow 0$. How
logarithmic corrections originated in irrelevant backscattering
and/or umklapp terms modify these results is an interesting
problem, subject of future work.

\begin{acknowledgments}
This work was partially supported by the Consejo Nacional de
Investigaciones Cient\'{\i}ficas y T\'ecnicas (CONICET) and
Universidad Nacional de La Plata (UNLP), Argentina. I am grateful
to Carlos Na\'on for useful comments and discussions, and for
encouraging me to write this article.
\end{acknowledgments}

\end{document}